\documentclass[preprint]{aastex}
\usepackage{graphicx}
\usepackage{epsfig}
\usepackage{amsmath}
\usepackage{lineno}
\usepackage{comment}
\usepackage{color}
\begin{document}
\shorttitle{Rotating Jets in Flux Emergence}
\shortauthors{Fang et al.}
\title{Rotating Solar Jets in Simulations of Flux Emergence with Thermal Conduction }

\author{Fang Fang, Yuhong Fan and Scott W. McIntosh}
\affil{High Altitude Observatory, National Center for Atmospheric Research, 
3090 Center Green Dr, Boulder, CO 80301}


\begin{abstract}
We study the formation of coronal jets through numerical simulation of the emergence 
of a twisted magnetic flux rope into a pre-existing open magnetic field. 
Reconnection inside the emerging flux rope in addition to that 
between the emerging and pre-existing fields give rise to 
the violent eruption studied. 
The simulated event closely resembles the coronal jets ubiquitously observed by 
Hinode/XRT and demonstrates that heated plasma is driven into the extended 
atmosphere above.  Thermal conduction implemented in the model allows us to 
qualitatively compare simulated and observed emission from such events. 
We find that untwisting field lines after the reconnection 
drive spinning outflows of plasma in the jet column. 
The Poynting flux in the simulated jet 
is dominated by the untwisting motions of the magnetic fields loaded with 
high-density plasma. The simulated jet is 
comprised of ``spires'' of untwisting field that are loaded with a mixture of 
cold and hot plasma and exhibit rotational motion of order 20~km/s and match 
contemporary observations.

\end{abstract}

\keywords{magnetohydrodynamics(MHD) --- Sun: corona --- Sun: activity}


\section{Introduction} \label{intro}
Coronal jets are hot columnar plasma ejecta that were first observed by 
the {\em Yohkoh} satellite \citep{shibata1992}. 
Subsequent (higher spatio-temporal resolution) observations with the {\em SOHO}, 
{\em STEREO} and {\em SDO} satellites have permitted study of these ubiquitous 
phenomena. Appearing as a collimated plasma ejecta from a low-lying loop-like 
structure they exhibit typical apparent speeds of $\sim$160~km/s, 
lifetimes of $\sim$10~minutes, heights of $\sim$50~Mm and cross-sectional 
widths of $\sim$8~Mm\citep{shimojo1996, savcheva2007}. 
The generally accepted view is that these high-speed coronal jets result from 
the reconnection of newly emerging flux system with the ambient (background) field. 
This process converts the magnetic energy in the system into thermal and kinetic 
energy for the plasma, heating it in the vicinity of the reconnection site, 
and producing the observed high-speed, hot, outflow of the plasma along 
the magnetic field lines.

Numerical simulations have been widely used to explore the physics behind 
coronal jets. \cite{shibata1992b}, \cite{yokoyama1995} and \cite{takasao2013} 
have studied the acceleration of jets through two-dimensional simulations of 
flux emergence \-- those simulations revealed fast outflows after the reconnection 
took place that were commensurate with the observations. 
Recent simulations have focused on the transition from 
a ``standard'' to ``blowout'' jet  as characterized by \citet{moore2010} by means
of emerging a twisted flux rope into an open field \citep{archontis2013}. 
Meantime, \cite{moreno-insertis2013} studied the onset mechanisms of 
the violent recurrent eruptions in the emerging region with a combination of
tether-cutting reconnection, kink instability, and mini-CME eruptions. 
Both later studies produced eruptive jets with dynamic properties that were more 
broadly comparable with the observational studies. 
Here, in order to study the emission structure of the ejected plasma column, 
we setup experiments with 
field-aligned heat conduction which produces a more realistic distribution of 
the thermal energy along the field lines during the reconnection event and the
jet. 


\section{Simulation Setup} \label{method}


We solve the ideal MHD equations with the implementation of heat conduction along 
the magnetic field lines within the Block Adaptive Tree Solar-wind Roe Upwind 
Scheme (BATS-R-US) that was developed at the University of Michigan 
\citep{powell1999,toth2012}:
\begin{equation}
  \frac{ \partial \rho}{\partial t} + \nabla \cdot (\rho {\bf u}) = 0, 
\end{equation}
\begin{equation}
  \label{momentum}
  \frac{\partial(\rho{\bf u})}{\partial t}+\nabla \cdot \left[\rho{\bf u}  
    {\bf u}+\left(p+\frac{{\bf B}{\bf B}}{8\pi}\right)\mathbf{I}-\frac{{\bf
        B}{\bf B}}{4\pi}\right]
  = \rho{\bf g},
\end{equation}
\begin{equation}
  \label{energy}
  \frac{\partial E}{\partial t} + \nabla \cdot \left[\left(                
    E + p + \frac{{\bf B}\cdot{\bf B}}{8\pi}\right){\bf u} - \frac{
      ({\bf u}\cdot{\bf B}){\bf B}}{4\pi}\right]
  = \frac{{\bf B}\cdot\nabla\left(\kappa_\parallel{\bf B}\cdot\nabla T\right)}{B^2},
\end{equation}
\begin{equation}
  \label{induction}
  \frac{\partial {\bf B} }{\partial t} = \nabla\times ({\bf u}\times{\bf B} ),
\end{equation} 
where $\rho$, ${\bf u}$, $E$, $p$, $T$, ${\bf B}$ and ${\bf g}$ are the density, 
velocity, total energy density, pressure, temperature, magnetic field and 
gravitational acceleration, respectively. $\kappa_\parallel$ denotes the 
coefficient of the thermal conductivity, with $\kappa_\parallel = \kappa_0T^{5/2}$
if $T > T_C = 3\times10^5 K$ and  $\kappa_\parallel = \kappa_0T_C^{5/2}$ if 
$T <= T_c$ \citep{linker2001}, 
where $\kappa_0 = 9.2\times10^{-12} W/(m^{-1}K^{-7/2})$, is the 
Spitzer conductivity. 
The correction of heat conduction coefficient at the transition-region 
temperatures allows us to expand the transition region artificially, in order 
to avoid the necessity of resolving the transition-region scale height of 1~km. 

We carry out two simulations here, one with heat conduction and the other without. 
The domain of our simulation covers 49.5(X)$\times$49.5(Y)$\times$79.2(Z) Mm$^3$, 
extending 4.9~Mm below the photosphere and 74.3~Mm into the corona with 
an initial background stratification of atmosphere in hydrostatic equilibrium,
following \cite{fan2001}. 
We then impose a uniform ambient magnetic field 
${\bf B} = B_0(-\sin\alpha {\bf e}_y + \cos\alpha {\bf e}_z)$, 
where $\alpha$ = 25$^{\circ}$ \citep{moreno-insertis2013}. 
Within the ambient field, we insert 
a horizontal flux rope in the {\bf x} direction at Z = -1.7~Mm, 
following \citep{fan2001} by ${\bf B} = B_{0}e^{-r^{2}/a^{2}}\hat{\boldsymbol x} + 
qrB_{0}e^{-r^{2}/a^{2}}\hat{\boldsymbol\theta}$, where $B_{0} = 3.8$~kG, 
is the strength of the magnetic fields at the central axis of the rope, 
$q = - 0.96/a$, is the rate of twist per unit length, 
and $a = 0.41$~Mm is the radius of Gaussian decay. 
The initial rope is density-depleted in the central portion, 
and the thermal pressure inside the flux rope is modified to maintain 
the total pressure balance. Panel (a) of Fig.~\ref{current} shows the 3D structure 
of the initial flux rope placed closely to the photosphere (represented by the 
purple line). The direction of magnetic fields in the flux rope and ambient fields 
are initialized to be anti-parallel at the interface. After the initialization, 
the buoyant, kink-stable flux rope emerges at its middle section and interacts 
with the ambient (open) field.


\section{Results} \label{result}


The initially-buoyant central part of the flux rope starts to rise immediately 
after being inserted into the convection zone, 
protrudes into the photosphere and expands further into the corona. 
Given the initial setup of the ambient field direction, 
the two flux systems are anti-parallel when they come into contact. 
This gives rise to the formation of a strong current sheet at the interface, 
as outlined by the purple rectangle in Panel (b) of Fig.~\ref{current}. 
The continuous emergence of the flux rope leads to thinning of the current sheet 
which eventually results in reconnection, releasing the free energy 
in the current sheet. The reconnection heats up the plasma, 
which is accelerated outward by the magnetic tension of the reconnected field. 
The emitted outflow has been categorized as a ``standard jet'' 
by \cite{moore2010}. 
In addition to the interface current between the ambient fields and the anchored
outer fields of the emerging flux rope, a strong current system develops in the 
central part 
of the emerging twisted flux rope \-- similar to the formation of current in other 
simulations without ambient fields \citep{manchester2004}. The central core field
of the flux rope, 
continues to rise and expand, forming another current sheet between the rising core
and ambient fields. In Panel (c) of Fig.~\ref{current} the core current is 
represented by the black rectangle while that between the core and ambient field 
is purple. Reconnections take place in these areas of strong current. 
Underneath the core, the reconnection lifts up the core fields, as shown by 
the field lines in the black rectangle. At the interface, the reconnection opens 
up the anchored emerging field, with a portion of the emerging field being 
redirected upward, at speeds up to 180~km/s. 
The other portion of the emerging flux system collapses 
into the lower atmosphere, forming a set of post-eruption loops (see Panel (c) 
in Fig.~\ref{mass}). 
This process, 
jet ejection resulting from reconnection at the base of the erupting core and 
the interface of the two flux systems, was dubbed a ``blowout jet'' by 
\cite{moore2010}.

\subsection{Mass ejection driven by the untwisting motion of the magnetic fields} 
\label{twistmotion}

Frequently the observed  X-ray and EUV jets undulate 
\citep[as originally noted by][]{shibata1992}. 
This property was explained as the movement of jets along helically twisted 
magnetic field lines that are accelerated by the $\mathbf{J}\times\mathbf{B}$ 
force during the relaxation and propagation of the magnetic twist along 
the field lines. \cite{canfield1996} also observed the spinning of H$\alpha$ surges 
when ejected outward using {\em Yohkoh} data, and found that the spin is uniform 
along the surges and is in such a direction that is consistent with the relaxation 
of the magnetic twist stored in the emerging fields. 
The untwisting motion found both in jets and H$\alpha$ surges 
\citep{cirtain2007,schmieder2013,moore2013} is interpreted as the propagation of 
MHD or torsional Alfv$\acute{\mathrm{e}}$n waves transporting the stored twist 
and helicity in the closed magnetic fields into the open ambient field. 
\cite{savcheva2007} carried out a statistical study of the properties of 
the transverse motion of the X-ray jets, and found speeds ranging from of 
0 to 35~km/s.

We can examine the transverse motion in our simulated jet by analyzing 
the velocity and current structures. When reconnection takes place between 
the emerging twisted magnetic flux rope and the (uniform) open ambient fields,
 a new set of open field lines forms at the interface and the plasma has 
an outward speed (u$_z$). The 3D structure of the reconnected field lines is 
illustrated in Panel (d)\--(f) of Fig.~\ref{current} where the lines are colored 
to show $|{\bf J}\cdot {\bf B}|/|{\bf B}|^2$ which measures the magnetic twist (d), 
u$_z$ (e) and u$_x$ (f) respectively. The new open field, differs from 
the initial ambient open field in that they inherits magnetic twist 
from the erupting twisted flux rope during the reconnection, as shown by 
the braidinging rods identified with black arrows. 
Shedding magnetic twist onto the new field yields an uneven 
distribution of the magnetic twist along the field lines, as shown by 
the $|{\bf J}\cdot {\bf B}|/|{\bf B}|^2$ color in Panel (d) of Fig.~ \ref{current}. 
Panel (e) shows that both upward and downward vertical motions are present 
in the jet column. 
Interestingly, on comparing panels (d) and (e) we see that the upflowing plasma 
is associated with field lines of strong twist while the downflowing plasma 
(as outlined by the black dashed rectangles) is 
associated with much weaker twist. Meanwhile, panel (f) shows the u$_x$ of the 
field lines, which is the component of velocity perpendicular to the plane. 
The reversal of u$_x$ indicates that the ejecta is rotating in the more twisted 
field lines (shown in Panel (d)). 
In contrast, in the dashed rectangle in panels (d)\--(f),
the field lines that were produced by the reconnection during 
the standard jet phase (see panel (b) of Fig.~\ref{current}) have already 
unwound themselves and possess much weaker magnetic twist (see panel (d)). 
On these field lines the plasma 
motion consists of a simple downward flow with a speed up to 66~km/s 
(see panel (e)), which corresponds to the speed under the combined effect of 
pressure gradient and gravity, 
with no apparent rotation (see panel (f)). The rotating upflow and 
non-rotating downflow of the plasma suggests that on the field lines with weak 
twist, the plasma falls back along the field lines under the effect of 
pressure gradient and gravitational acceleration. 
These are consistent with observations by \cite{liuw2009}, 
where they reported untwisting motion with transverse oscillation 
velocities ranging from 26 to 151 km/s in the outward flow 
while the downward component exhibits no transverse motion.

\begin{figure*}[ht!]
  \begin{minipage}[t] {1.0\linewidth}
    \begin{center}
      \includegraphics[width=150mm]{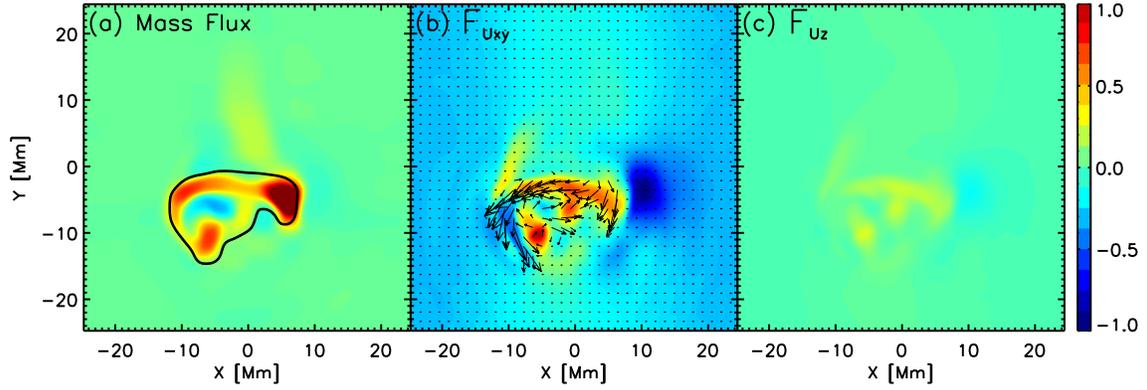}
    \end{center}
  \end{minipage}\hfill
  \caption{Mass flux density in unit of $2\times10^{-7}~kg/s/m^2$ (a), 
    Poynting flux density associated with horizontal velocity (b) 
    and vertical velocity (c) in unit of $3\times10^3~J/s/m^2$ 
    on the Z = 30 Mm plane at t = 59 min. 
    Black line represents the contour of $\rho/\rho_{0}$ = 1.5, 
    where  $\rho_{0}$ is the initial density value at Z = 30 Mm. 
    Arrows in Panel (b) show the horizontal velocity u$_{xy}$.}
  \label{flux}
\end{figure*}

\begin{figure*}[ht!]
  \begin{minipage}[t] {1.0\linewidth}
    \begin{center}
      \includegraphics[width=130mm]{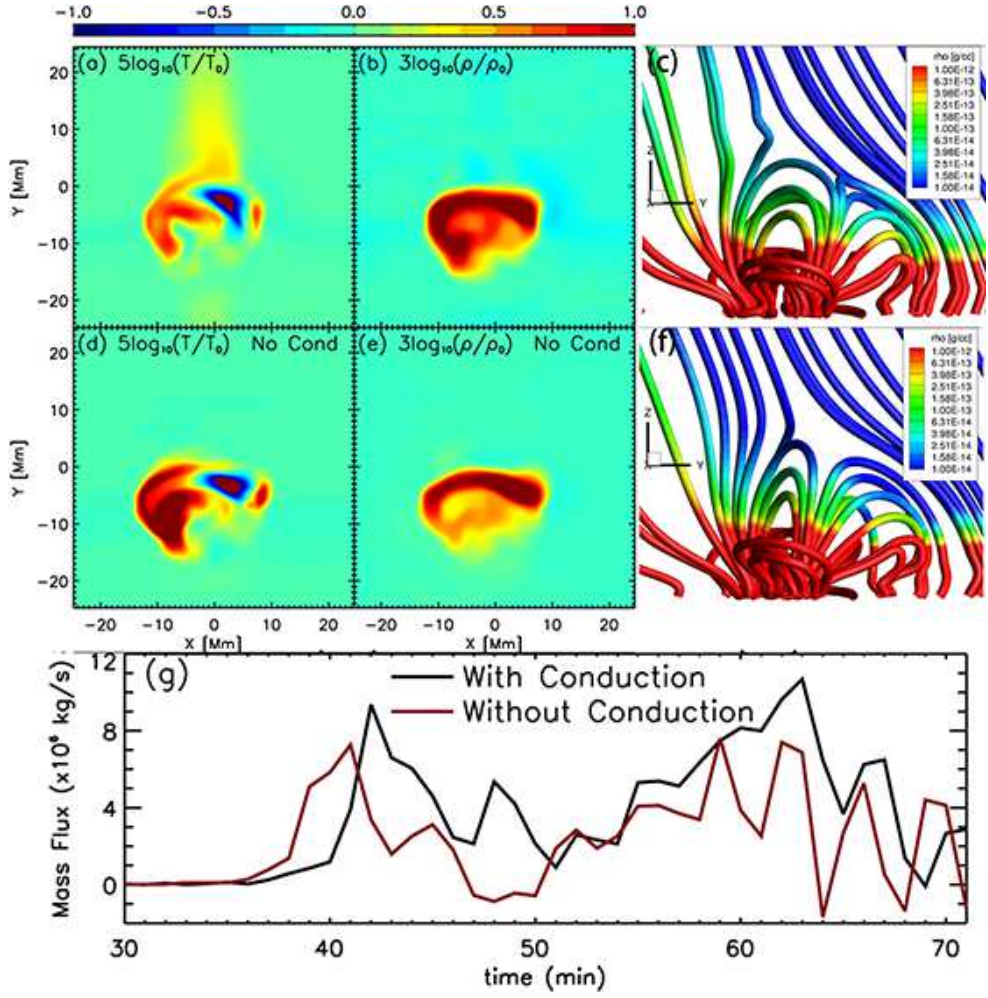}
    \end{center}
  \end{minipage}\hfill
  \caption{(a and b) 5$\log_{10}(T/T_0)$ (a) and 3$\log_{10}(\rho/rho_0)$ (b)
    at t = 59 min on Z = 30 Mm. (d and e) the same but from comparison run without
    heat conduction. (c and f) 3D magnetic field lines colored by $\rho$ in the
    closed-loop region at t = 40 min with (c) and without (f) heat conduction. 
    (g) mass flux across Z=30 Mm during simulations with (black line) and without 
    (red) heat conduction.}
  \label{mass}
\end{figure*}

To study the relaxation of the reconnected field lines we examine the mass and 
energy transport in the corona during the eruption. We calculate the Poynting flux 
density associated with horizontal and vertical velocities at the plane Z = 30 Mm by:
\begin{eqnarray}
  F_{u_{xy}}& = &- \frac{1}{8\pi}\left(B_{x}u_{\perp x} + B_{y}u_{\perp y}\right)B_{z} ,\\
  F_{u_z} & = &\frac{1}{8\pi}\left(B_{x}^{2} + B_{y}^{2}\right)u_{\perp z},
\end{eqnarray}
where u$_{\perp x}$, u$_{\perp y}$ and u$_{\perp z}$ are the x-, y- and z-components 
of the velocity perpendicular to the magnetic field. 
Fig.~\ref{flux} shows the mass flux density and Poynting flux 
density with horizontal and vertical flows at t = 59 min, with 
arrows in the panels (b) representing the horizontal velocity u$_{xy}$. 
We find that the horizontal velocity clearly presents a pattern of rotating motion. 
Comparing the Poynting flux component F$_{u_{xy}}$ and F$_{u_z}$ 
shows that it is the horizontal untwisting motion 
that dominates the energy flux into the corona. 
In addition, the mass transported into the corona consists of high-density plasma, 
shown by the contour lines of $1.5\rho_0$ in panel (a), with mixed 
hot and cool temperatures \-- panel (a) in Fig.~\ref{mass}. 
And the mass ejection is accompanied by the Poynting energy flux associated 
with the horizontal untwisting motion, shown in panel (b) of Fig.~\ref{flux}.

The mass ejection into the corona is strongly facilitated by the field-aligned
thermal conduction, as shown in Fig.~\ref{mass} comparing results from 
simulations with and without conduction. We find that 
the mass increase (3$\log_{10}(\rho/\rho_0)$) at t = 59 min in the blowout jet 
is stronger with thermal conduction (Panel (b)). 
Also noticeable is that the temperature structure in the 
non-conduction run (Panel (d)) consists of more extreme components as compared to 
the case with conduction (Panel (a)). 
Panels (c) and (f) show the 3D structures of the post-eruption loops 
after the reconnection at t = 40 min, colored by the density.  
In these post-eruption loops, heat conduction redistributed the thermal energy 
from the reconnection site along the field lines, which drives more plasma 
upward into the corona. This upward mass transport gives rise to the observed 
stronger mass loading in the closed loops in the simulation with heat conduction 
(panel (c)).
In panel (g), we calculate the mass flux across Z = 30 Mm plane in the corona, 
and find that multiple mass ejections occur during both simulations, 
with the first peak of mass flux (occurring  at time t = 41 and 42 min  
for non-conduction and conduction run, respectively) 
associated with standard jet. Clearly, heat conduction greatly increases the 
mass ejection in the standard phase, from 1.46$\times$10$^9$ in the non-conduction 
run to 2.30$\times$10$^9$ kg in the conduction one. 
The total ejection of mass into the corona during the 71 min of simulation 
is 9.2$\times$10$^9$ kg in the case with heat conduction, as compared with only 
5.8$\times$10$^9$ kg in the non-conduction one with more instances of negative 
mass flux.


\subsection{Synthetic Emission}


Mass ejected into the corona is guided outward by the magnetic field lines 
in simulations with field-aligned thermal conduction.
To evaluate the observational effects of the mass ejection, we calculate
the coronal and upper transition-region (\ion{Fe}{8} dominated \-- 171\AA) 
and transition-region 
(\ion{He}{2} dominated \-- 304\AA) synthetic emission using the 
Atmospheric Imaging Assembly (AIA) temperature response functions. 
Fig.~\ref{emission} shows the emission at different times during the eruption, 
as viewed along the axial direction of the flux rope. 
In the coronal line \-- panels (a) and (c) \-- two jet ``spires'' form 
inside the column, one (Panel (a)) related to the reconnection 
between the ambient fields 
with the fields in the outer periphery of the emerging flux rope and another 
(Panel (c))
which forms when the core fields rise into the corona and reconnects with 
the open field. The apprearance of the second spire (Panel (c)) starts 
after the cool transition-region plasma is ejected into the lower corona, 
as outlined in the dashed square in Panel (b), and clearly at transition-region 
temperature shown in Panel (e). The jet column widens and brightens in 
both channels (panels (c) and (f)), grossly resembling the blowout jets 
reported in observations \citep{moore2010}. 
In contrast, emission from the non-conduction simulation 
(panels (g-i)) clearly show that most of the massive material 
ejected during the eruption, outlined by the dashed square in Panel (h), 
cannot be levitated beyond the lower corona, and falls back after the eruption
(see also the online movies). 
Interestingly, the post-eruption loops (green rectangles in Fig.~\ref{emission}), 
which are closed field lines resulting from the reconnection 
of the ambient fields and that in the emerging flux rope, appears brighter 
with heat conduction, 
both at t = 40 min during the standard phase (panels (a) and (g)), and at
t = 55 min during the blowout jet (panels (c) and (i)). 

In Fig.~\ref{doppler} we calculate the 304\AA{} 
intensity-weighted line-of-sight (LOS, x-direction) velocity V$_{LOS}$ and 
we see the 
reversal of sign in V$_{LOS}$ during the eruption which suggests that the jet column 
is rotating at a speed of $\sim$20~km/s that is consistent with observations 
\citep{liuw2009}. 
The swirling spire in panel (d) corresponds to braided magnetic fields loaded with
dense plasma. 
The dashed rectangle in panels (a) through (d) 
identifies another ejection of cool plasma (panel (b)) in the blowout jet that 
is followed by the brightening and widening of the jet in the corona, 
which corresponds to the 
peak in mass flux at t = 63 min in Panel (g) of Fig.~\ref{mass}. 
At the corresponding times, the non-conduction simulation produces 
a similar reversal pattern in V$_{LOS}$, seen in Panel (e) and (f). 
However, the mass loading does not delineate the field lines as well in this case 
without the inclusion of thermal conduction. 
Hence, the structure of intensity-weighted V$_{LOS}$ does not resemble 
the configuration of the magnetic fields.


\section{Summary and Conclusions} \label{conclusion}

\begin{figure*}[ht!]
  \begin{minipage}[t] {1.0\linewidth}
    \begin{center}
      \includegraphics[width=150mm]{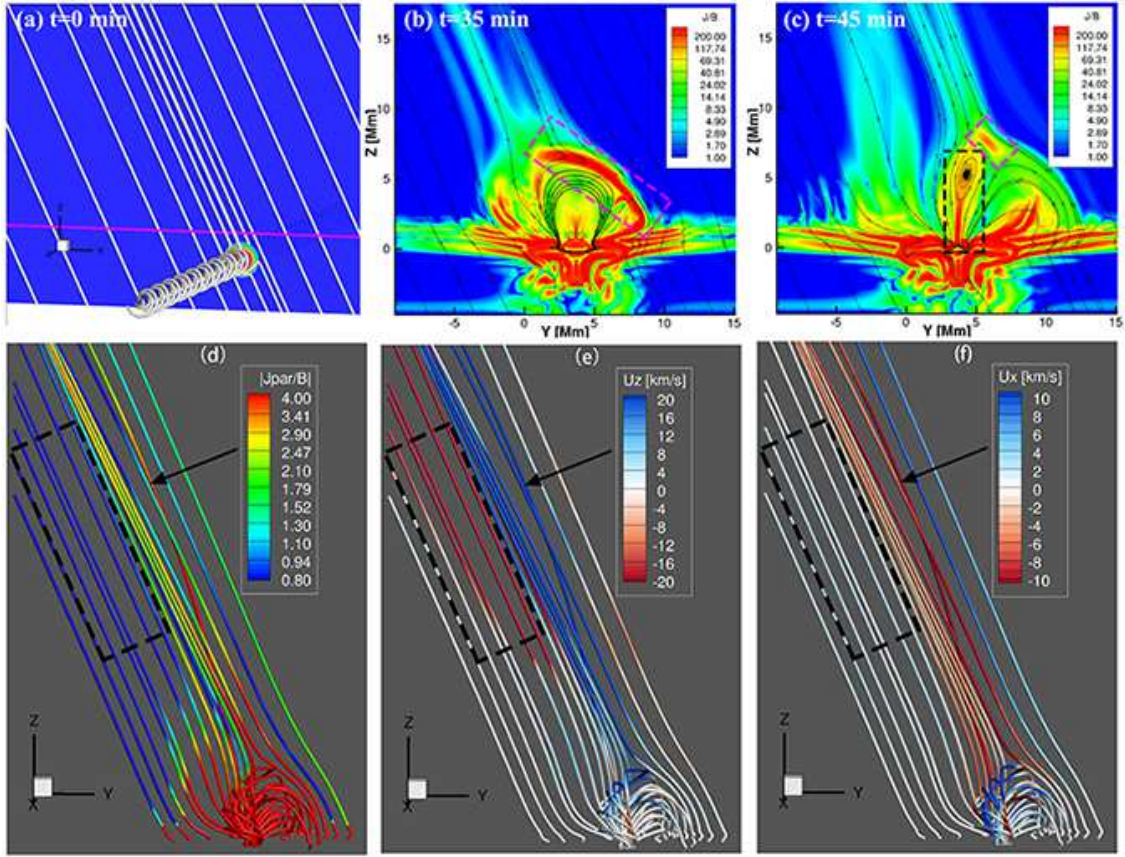}
    \end{center}
  \end{minipage}\hfill
  \caption{(a) 3D structure of the initial magnetic fields with X=0 Mm plane. 
    The purple line represents the location of the 
    photosphere at Z = 0~Mm.
    (b and c) $|J|/|B|$ on the X = 0~Mm plane with black lines showing the 
    magnetic fields in the plane at t = 35 (b) and 45 (c) min. The dashed
    rectangles outline regions of strong current sheet. 
    (d-f) 3D structure of the magnetic fields colored by 
    $|{\bf J}\cdot {\bf B}|/|{\bf B}|^2$ (d), u$_z$ (e) and u$_x$ (f)
    at t = 51 min. 
    The arrows point to the unwinding magnetic field lines and 
    rectangles outline areas of downflowing plasma with no apparent rotating motion
    along magnetic field lines with low twist.}
  \label{current}
\end{figure*}

\begin{figure*}[ht!]
  \begin{minipage}[t] {1.0\linewidth}
    \begin{center}
      \includegraphics[width=130mm]{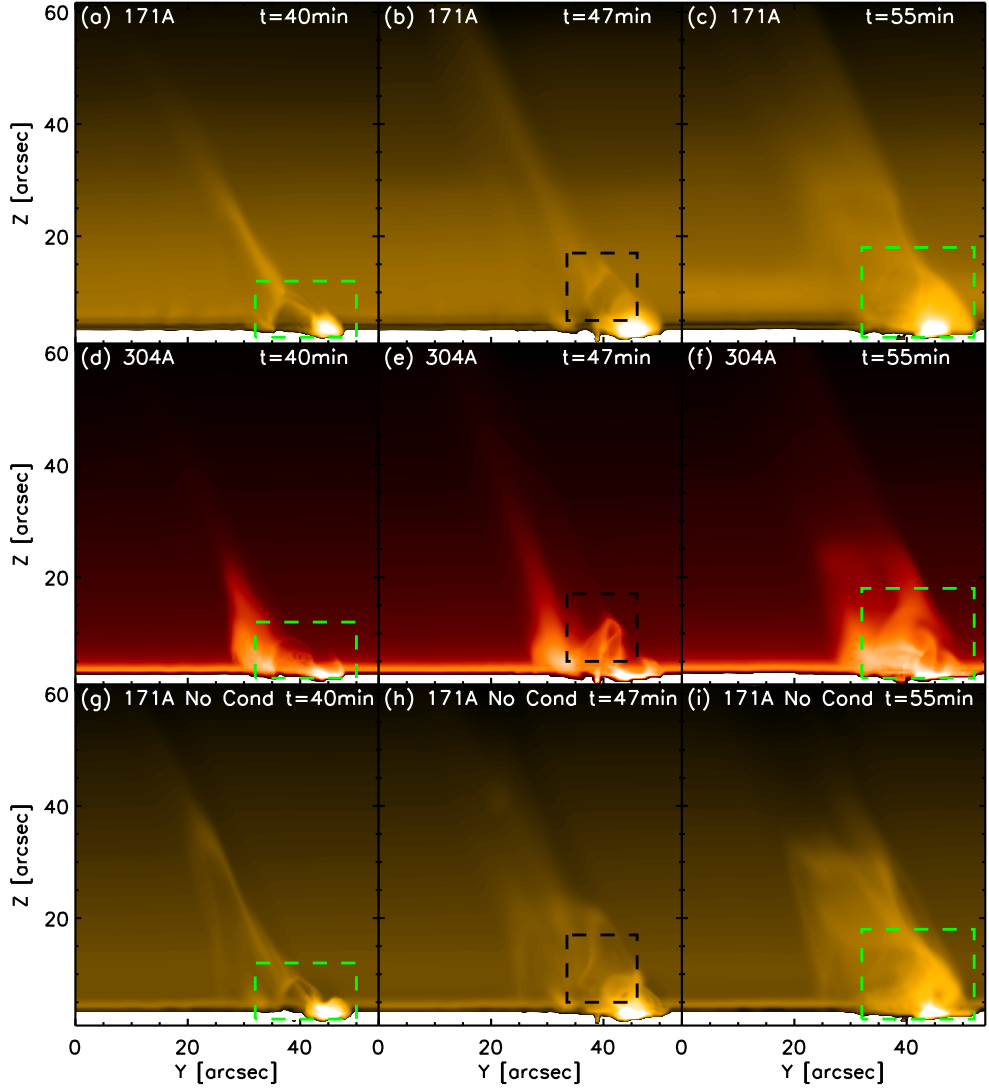}
    \end{center}
  \end{minipage}\hfill
  \caption{Synthetic emission in the AIA 171\AA{} (a-c,g-i) and 304\AA{} (d-f) 
    channels at t = 40, 47 and 55 min from simultaions with (a-f) and 
    without (g-i) thermal conduction. 
    The green rectangles show the post-reconnection loops. 
    The dashed black square outlines region of mass ejection into the corona. 
    Movies showing the emission at varying wavelengths comparing 
    the two simulations are available in the electronic version. }
  \label{emission}
\end{figure*}

\begin{figure*}[ht!]
  \begin{minipage}[t] {1.0\linewidth}
    \begin{center}
      \includegraphics[width=130mm]{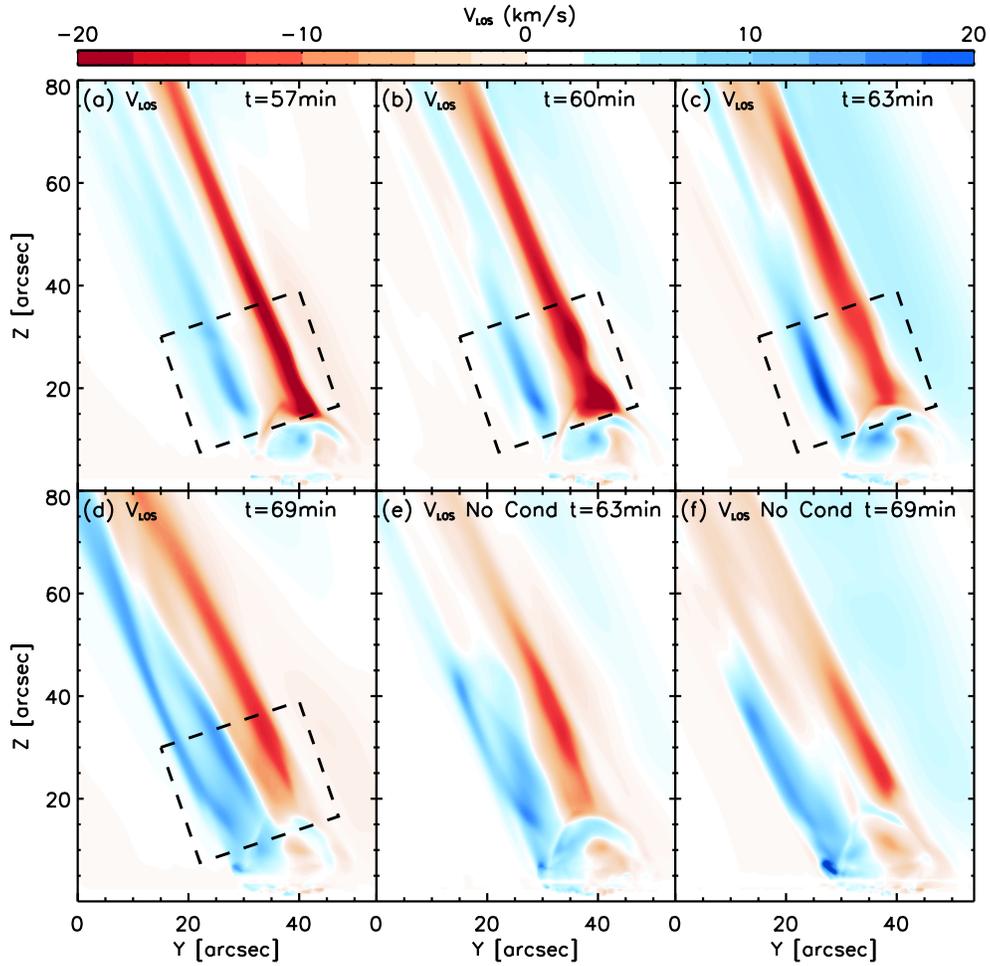}
    \end{center}
  \end{minipage}\hfill
  \caption{The AIA 304\AA{} intensity-weighted LOS velocity during the eruption
    from simulations with (a-d) and without (e-f) heat conduction. 
    The dashed rectangle outlines the region of mass ejection into the corona. 
    Movies showing the intensity-weighted LOS velocity at varying
    wavelengths comparing the two simulations are available in the electronic 
    version.}
  \label{doppler}
\end{figure*}

We present numerical simulations of coronal jets which occur during the emergence 
of a twisted flux rope into an ambient open field. The emergence and expansion 
of the flux rope into the ambient field gives rise to magnetic reconnection at 
multiple locations during the phases of the eruption 
\citep{archontis2013, moreno-insertis2013}. 
We find that a mixture of hot and cold plasma is driven upward along the unwinding 
field lines during the eruption which produces the observed oscillatory transverse 
motion of the ejected jet spires, while the companion downward motion in 
the jet exhibits no rotation \citep{liuw2009}. The rotating upflowing motion and 
the simple downflow are correlated with the spatial distribution of magnetic twist 
in the jet column, with stronger magnetic twist in the rotating upflowing 
plasma (see Fig.~\ref{current}). This strongly suggests that material is accelerated 
upward by the Lorentz force. Further, the rotating motion of the ejected plasm has 
the same sense as the unwinding reconnected field lines. Analysis of the energy flux 
in the system shows that the Poynting flux in the corona, dominated by 
the rotating motion of the structure, 
coincides with the outward mass transport of dense plasma. 
During the blowout jet, multiple ejecta take place low in the atmosphere, 
propagating along the twisted magnetic field lines after the reconnection 
between the rising core fields of the flux rope and the ambient fields, 
drive plasma of chromospheric and transition-region temperatures up into 
the corona by Lorentz force in addition to the hot reconnection outflow. 
Field-aligned thermal conduction efficiently distributes the energy release from 
the reconnection region into the lower atmopshere and promotes the ejection of dense 
plasma into the corona along the field lines, 
which is essential for comparing synthetic emission 
(Fig. \ref{emission} and \ref{doppler}) with observations. 
Without conduction, the ejected mass does not delineate field lines as well, 
and more importantly, falls back to the lower atmosphere due to
lack of support, resulting in a reduced total mass ejection into the corona. 
The total mass in 
the corona is increased by a factor of 2$\%$ during the blowout jets. 
These jets with hot mass outflow into the corona, 
have been suggested as a source of heated mass of the solar wind 
\citep{depontieu2007,mcintosh2010}, will be studied in more detail in the future. 

\acknowledgments
We thank the referee for valuable comments in improving the manuscript. 
F.~F. is supported by Advanced Study Program of the National Center for 
Atmospheric Research which is sponsored by the National Science Foundation. 
The simulations described here were carried out on the Yellowstone Supercomputer 
at NCAR and Stampede system in the Texas Advanced Computing Center (TACC)
at the University of Taxas at Austin. 

\clearpage

\begin{thebibliography}{19}
\expandafter\ifx\csname natexlab\endcsname\relax\def\natexlab#1{#1}\fi

\bibitem[{{Archontis} \& {Hood}(2013)}]{archontis2013}
{Archontis}, V. \& {Hood}, A.~W. 2013, \apjl, 769, L21

\bibitem[{{Canfield} {et~al.}(1996){Canfield}, {Reardon}, {Leka}, {Shibata},
  {Yokoyama}, \& {Shimojo}}]{canfield1996}
{Canfield}, R.~C., {Reardon}, K.~P., {Leka}, K.~D., {Shibata}, K., {Yokoyama},
  T., \& {Shimojo}, M. 1996, \apj, 464, 1016

\bibitem[{{Cirtain} {et~al.}(2007){Cirtain}, {Golub}, {Lundquist}, {van
  Ballegooijen}, {Savcheva}, {Shimojo}, {DeLuca}, {Tsuneta}, {Sakao}, {Reeves},
  {Weber}, {Kano}, {Narukage}, \& {Shibasaki}}]{cirtain2007}
{Cirtain}, J.~W., {Golub}, L., {Lundquist}, L., {van Ballegooijen}, A.,
  {Savcheva}, A., {Shimojo}, M., {DeLuca}, E., {Tsuneta}, S., {Sakao}, T.,
  {Reeves}, K., {Weber}, M., {Kano}, R., {Narukage}, N., \& {Shibasaki}, K.
  2007, Science, 318, 1580

\bibitem[{{De Pontieu} {et~al.}(2007){De Pontieu}, {McIntosh}, {Carlsson},
  {Hansteen}, {Tarbell}, {Schrijver}, {Title}, {Shine}, {Tsuneta}, {Katsukawa},
  {Ichimoto}, {Suematsu}, {Shimizu}, \& {Nagata}}]{depontieu2007}
{De Pontieu}, B., {McIntosh}, S.~W., {Carlsson}, M., {Hansteen}, V.~H.,
  {Tarbell}, T.~D., {Schrijver}, C.~J., {Title}, A.~M., {Shine}, R.~A.,
  {Tsuneta}, S., {Katsukawa}, Y., {Ichimoto}, K., {Suematsu}, Y., {Shimizu},
  T., \& {Nagata}, S. 2007, Science, 318, 1574

\bibitem[{{Fan}(2001)}]{fan2001}
{Fan}, Y. 2001, \apjl, 554, L111

\bibitem[{{Linker} {et~al.}(2001){Linker}, {Lionello}, {Miki{\'c}}, \&
  {Amari}}]{linker2001}
{Linker}, J.~A., {Lionello}, R., {Miki{\'c}}, Z., \& {Amari}, T. 2001, \jgr,
  106, 25165

\bibitem[{{Liu} {et~al.}(2009){Liu}, {Berger}, {Title}, \&
  {Tarbell}}]{liuw2009}
{Liu}, W., {Berger}, T.~E., {Title}, A.~M., \& {Tarbell}, T.~D. 2009, \apjl,
  707, L37

\bibitem[{{Manchester} {et~al.}(2004){Manchester}, {Gombosi}, {DeZeeuw}, \&
  {Fan}}]{manchester2004}
{Manchester}, IV, W., {Gombosi}, T., {DeZeeuw}, D., \& {Fan}, Y. 2004, \apj,
  610, 588

\bibitem[{{McIntosh} {et~al.}(2010){McIntosh}, {Innes}, {de Pontieu}, \&
  {Leamon}}]{mcintosh2010}
{McIntosh}, S.~W., {Innes}, D.~E., {de Pontieu}, B., \& {Leamon}, R.~J. 2010,
  \aap, 510, L2

\bibitem[{{Moore} {et~al.}(2010){Moore}, {Cirtain}, {Sterling}, \&
  {Falconer}}]{moore2010}
{Moore}, R.~L., {Cirtain}, J.~W., {Sterling}, A.~C., \& {Falconer}, D.~A. 2010,
  \apj, 720, 757

\bibitem[{{Moore} {et~al.}(2013){Moore}, {Sterling}, {Falconer}, \&
  {Robe}}]{moore2013}
{Moore}, R.~L., {Sterling}, A.~C., {Falconer}, D.~A., \& {Robe}, D. 2013, \apj,
  769, 134

\bibitem[{{Moreno-Insertis} \& {Galsgaard}(2013)}]{moreno-insertis2013}
{Moreno-Insertis}, F. \& {Galsgaard}, K. 2013, \apj, 771, 20

\bibitem[{{Powell} {et~al.}(1999){Powell}, {Roe}, {Linde}, {Gombosi}, \& {de
  Zeeuw}}]{powell1999}
{Powell}, K.~G., {Roe}, P.~L., {Linde}, T.~J., {Gombosi}, T.~I., \& {de Zeeuw},
  D.~L. 1999, Journal of Computational Physics, 154, 284

\bibitem[{{Savcheva} {et~al.}(2007){Savcheva}, {Cirtain}, {Deluca},
  {Lundquist}, {Golub}, {Weber}, {Shimojo}, {Shibasaki}, {Sakao}, {Narukage},
  {Tsuneta}, \& {Kano}}]{savcheva2007}
{Savcheva}, A., {Cirtain}, J., {Deluca}, E.~E., {Lundquist}, L.~L., {Golub},
  L., {Weber}, M., {Shimojo}, M., {Shibasaki}, K., {Sakao}, T., {Narukage}, N.,
  {Tsuneta}, S., \& {Kano}, R. 2007, \pasj, 59, 771

\bibitem[{{Schmieder} {et~al.}(2013){Schmieder}, {Guo}, {Moreno-Insertis},
  {Aulanier}, {Yelles Chaouche}, {Nishizuka}, {Harra}, {Thalmann}, {Vargas
  Dominguez}, \& {Liu}}]{schmieder2013}
{Schmieder}, B., {Guo}, Y., {Moreno-Insertis}, F., {Aulanier}, G., {Yelles
  Chaouche}, L., {Nishizuka}, N., {Harra}, L.~K., {Thalmann}, J.~K., {Vargas
  Dominguez}, S., \& {Liu}, Y. 2013, \aap, 559, A1

\bibitem[{{Shibata} {et~al.}(1992{\natexlab{a}}){Shibata}, {Ishido}, {Acton},
  {Strong}, {Hirayama}, {Uchida}, {McAllister}, {Matsumoto}, {Tsuneta},
  {Shimizu}, {Hara}, {Sakurai}, {Ichimoto}, {Nishino}, \&
  {Ogawara}}]{shibata1992}
{Shibata}, K., {Ishido}, Y., {Acton}, L.~W., {Strong}, K.~T., {Hirayama}, T.,
  {Uchida}, Y., {McAllister}, A.~H., {Matsumoto}, R., {Tsuneta}, S., {Shimizu},
  T., {Hara}, H., {Sakurai}, T., {Ichimoto}, K., {Nishino}, Y., \& {Ogawara},
  Y. 1992{\natexlab{a}}, \pasj, 44, L173

\bibitem[{{Shibata} {et~al.}(1992{\natexlab{b}}){Shibata}, {Nozawa}, \&
  {Matsumoto}}]{shibata1992b}
{Shibata}, K., {Nozawa}, S., \& {Matsumoto}, R. 1992{\natexlab{b}}, \pasj, 44,
  265

\bibitem[{{Shimojo} {et~al.}(1996){Shimojo}, {Hashimoto}, {Shibata},
  {Hirayama}, {Hudson}, \& {Acton}}]{shimojo1996}
{Shimojo}, M., {Hashimoto}, S., {Shibata}, K., {Hirayama}, T., {Hudson}, H.~S.,
  \& {Acton}, L.~W. 1996, \pasj, 48, 123

\bibitem[{{Takasao} {et~al.}(2013){Takasao}, {Isobe}, \&
  {Shibata}}]{takasao2013}
{Takasao}, S., {Isobe}, H., \& {Shibata}, K. 2013, \pasj, 65, 62

\bibitem[{{T{\'o}th} {et~al.}(2012){T{\'o}th}, {van der Holst}, {Sokolov}, {De
  Zeeuw}, {Gombosi}, {Fang}, {Manchester}, {Meng}, {Najib}, {Powell}, {Stout},
  {Glocer}, {Ma}, \& {Opher}}]{toth2012}
{T{\'o}th}, G., {van der Holst}, B., {Sokolov}, I.~V., {De Zeeuw}, D.~L.,
  {Gombosi}, T.~I., {Fang}, F., {Manchester}, W.~B., {Meng}, X., {Najib}, D.,
  {Powell}, K.~G., {Stout}, Q.~F., {Glocer}, A., {Ma}, Y.-J., \& {Opher}, M.
  2012, Journal of Computational Physics, 231, 870

\bibitem[{{Yokoyama} \& {Shibata}(1995)}]{yokoyama1995}
{Yokoyama}, T. \& {Shibata}, K. 1995, \nat, 375, 42

\end{thebibliography}


\clearpage

\end{document}